\RequirePackage{fix-cm}
\documentclass{svjour3}                     
\smartqed  
\usepackage{graphicx}
\begin{document}

\title{Stability analysis of VBT Echelle spectrograph for precise Radial Velocity measurements}
\titlerunning{Analyzing the stability of VBT Echelle Spectrograph}
\author{Sireesha Chamarthi,
       Ravinder K. Banyal,
       S. Sriram,  Gajendra Pandey
       }
\institute{Sireesha Chamarthi \at
              Indian Institute of Astrophysics, Bangalore \\
             \email{sireesha@iiap.res.in}
           \and
           Ravinder K. Banyal,  \at
           Indian Institute of Astrophysics, Bangalore\\
           \email{banyal@iiap.res.in}
           \and
            S. Sriram,  \at
           Indian Institute of Astrophysics, Bangalore\\
           \email{ssr@iiap.res.in}
           \and
           Gajendra Pandey,  \at
           Indian Institute of Astrophysics, Bangalore\\
           \email{pandey@iiap.res.in}
}
\date{Received: date / Accepted: date}
\maketitle
\begin{abstract}
A fiber-fed Echelle spectrograph at 2.3~m Vainu Bappu Telescope (VBT), Kavalur, has been in operation since 2005. Owing to various technological advancements in precision spectroscopy in recent years,  several avenues have been opened in observational astronomy. These developments have created a demand to improve the Doppler precision of our spectrograph. Currently, the stability of the instrument is compromised by the temperature and pressure fluctuations inside the Echelle room. Further, a better wavelength calibration approach is needed to carefully track and disentangle the instrumental effects from stellar spectra. While planning a possible upgrade with an Iodine absorption gas cell, we measured the raw stability of the spectrograph using a series of calibration frames taken with the Th-Ar gas discharge lamp. The time series data were analysed with cross-correlation method and the shift in  Th-Ar emission lines was accurately measured across different Echelle orders.  In this paper, we present our stability analysis methodology and results for the Kavalur spectrograph. We also identify possible sources of error and discuss our strategy to mitigate them.

\keywords{Spectrograph \and Stability Analysis \and Radial Velocity \and ThAr\and Iodine Absorption cell}
\end{abstract}

\section{Introduction}
\label{intro}
Accurate measurement of radial velocities (RV) has enabled astronomers to detect hundreds of extra-solar planets around other stars \cite{Intro_0}. Stellar oscillations measured with precision RV technique are used for probing the stellar interiors \cite{Intro_1}. Variability of fundamental constants is another important area of research which requires extended observations of minute Doppler shifts of distant sources, measured with extreme precision and accuracy \cite{Intro_2}. The Echelle spectrograph operating at VBT, Kavalur was designed to carry out general-purpose high-resolution observations in the visible band. However, with the emergence of new research areas, there is a renewed interest among the astronomical community to expand the scientific capability of the spectrograph. One particular area of interest is to utilise the spectrograph for precision RV measurements, pertaining to detection and characterization of extra-solar planets.

Instrument stability is one of the most critical requirements to achieve the desired RV precision.  Even the RV detection of short-period giant planets (hot-Jupiters), requires the Doppler precision at the level of $\sim$100~m/s. On the detector plane, an RV change of 100~m/s translates to a spectral line shift smaller than 1/10 the of the CCD pixel. In other words, in accordance with Doppler formula, $\Delta\lambda/\lambda=v/c$, a 100~m/s RV would shift a typical spectral line of the star at 650~nm by $\sim2\times 10^{-4}$~nm. Such tiny shifts are often below the stability limit of most general purpose spectrographs. Thus a small Doppler shift is therefore easily drowned in large instrumental drifts caused by randomly changing environmental conditions (temperature and pressure fluctuations) and illumination errors caused by telescope flexure and imperfect guiding. At most observatories, the nightly to weekly instrument drift can easily exceed several 100~m/s.

Another source of RV uncertainty is wavelength calibration errors. The path traversed by the star-light and that of the calibration lamp light is not the same. They illuminate different parts of the spectrograph, thus contributing to errors in wavelength solution. Often due to design constraints, the stellar spectra and calibration exposures are taken at different epochs. In such cases, the spectrograph drifts during the stellar observations remain unaccounted.

A very high RV precision ($\sim1$~m/s) is now achievable with highly stable and state-of-the-art spectrographs such as HARPS (High  Accuracy  Radial  velocity  Planet Searcher). This instrument operates under strictly regulated vacuum conditions and uses two fibres -one for Th-Ar lamp and another for star light, to simultaneously record the reference and stellar spectra respectively on CCD. A laser frequency comb providing an extremely stable grid of reference lines for wavelength calibration has also been tested on HARPS to extend RV search for Earth analogues orbiting around other stars. Following the successful design philosophy of HARPS, many new instruments have already been built along the same lines, while some are in advanced stages of completion.

The same is not the case for existing non-stabilized spectrographs. Apart from the need for upgrades to achieve a modest level of stability against ambient perturbations, a suitable observation and calibration strategy is necessary to track instrumental drift. To that effect, an approach developed by Marcy and Butler using an Iodine absorption cell, was successful \cite{Iodine_1}. It is now extensively used on many spectrographs worldwide. The Iodine gas cell when placed in the path of the star-light, superimposes dense absorption features of Iodine (within the 500-630~nm band) onto the stellar spectrum. The advantage of this technique is that the instrumental noise remains common to stellar as well as Iodine lines. By modelling the composite spectra of the star and Iodine, a true RV signal of the target star can be extracted out \cite{Iodine_1}.
\begin{figure}[ht]
\begin{center}
\begin{tabular}{c}
\includegraphics[scale=0.42]{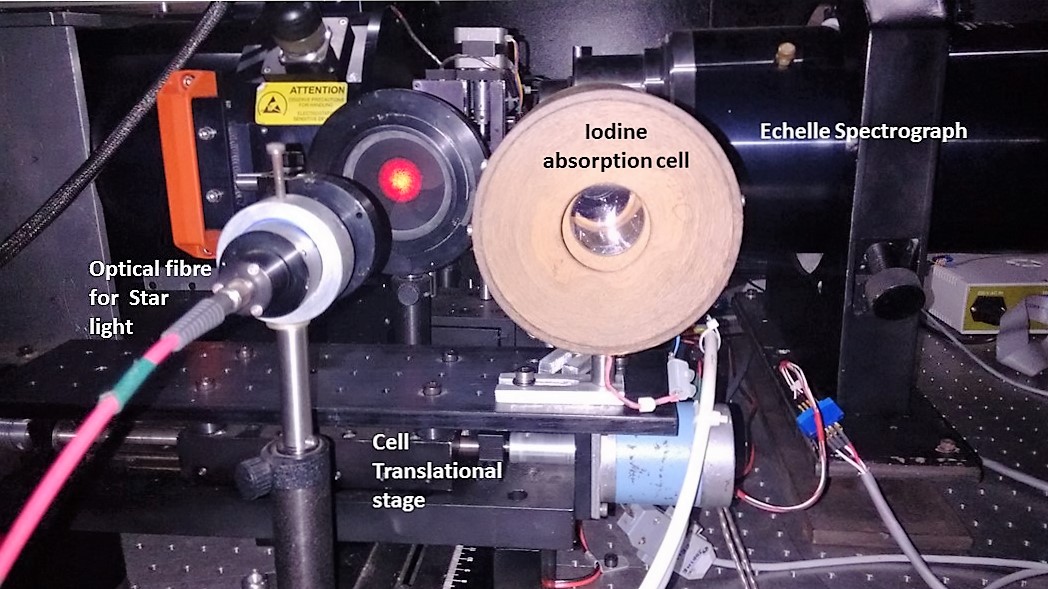}
\end{tabular}
\end{center}
\caption
{\label{fig:cell}
Echelle spectrograph assembly at the  VBT, Kavalur. Optical fibre seen in the foreground of the image transports stellar light and Th-Ar calibration light from prime focus of the telescope to the spectrograph room. An Iodine absorption cell (shown retracted) is under plan to facilitate simultaneous illumination of the spectrograph with stellar and reference light to improve the RV precision within 5000-6300~${\AA}$ band.}
\end{figure}
We  are  developing  a  thermally  stabilised  Iodine absorption cell for Kavalur Echelle. The cell will be placed
in front of the entrance slit of the spectrograph (as seen in Figure \ref{fig:cell}) in a collimated beam following the fibre. It is moved \textit{in} and \textit{out} of the beam path using a motorized translation stage. The Echelle spectrograph is housed inside an air-conditioned room with a temperature stability of $\pm1^{\circ}$C.

The inherent instability of the spectrograph needs to be carefully examined before developing a complete RV model with Iodine cell. A certain level of instrument stability must be ensured before attempting to improve the RV precision with Iodine technique. In this paper, we report our efforts towards understanding and improving the inherent stability of the spectrograph. A short description of the spectrograph and related hardware is given in Section 2. The methodology and procedure adopted for stability analysis are described in Section 3. The stability analysis of VBT Echelle is presented in Section 4. In Section 5 we have discussed the possible area of improvement and plan for upgradations. Our concluding remarks are presented in Section 6.

\section{Spectrograph Description}
A complete description and original design of the spectrograph is discussed in Ref \cite{VBT_2}. The light reflected from
the f/3 primary mirror of VBT is injected into a 100~$\mu$m  multi-mode optical fibre located at the prime focus of the
telescope. The fibre carries the stellar light to the spectrograph room where the beam exiting the fibre is collimated using a f/3 relay optics and is focused on the slit by a f/5 system. A collimator assembly inside the spectrograph
accepts the f/5 beam from the slit and directs the collimated
beam onto the cross-dispersing prism. The vertical dispersion of the light is attained with the help of the prism. The Echelle grating is set up for dispersion in the horizontal direction. This high dispersion spectrum is then fed-backward onto the cross-dispersing prism. The output of the prism illuminates the imaging detector through the collimator which acts as a camera in the second pass. The large blaze angle (${\theta}${\tiny B}= $70\,^{\circ}$) of the Echelle grating configured in Littrow mode, enables high resolution ($R=60,000$) stellar observations. This is supplemented with large wavelength coverage from 4000-10,000 {\AA}.The
specifications of the spectrograph components are listed in
Table 1. Th–Ar calibration lamp assembly placed at the
prime-cage of the telescope is used for wavelength calibration frames for science observations. The light from the lamp and the flat field source is fed to the spectrograph unit with the help of the same  100~$\mu$m optical fibre which is used to feed the starlight.
\begin{table}[ht]
\caption{Specifications of the spectrograph components.}
\label{tab:Echelle}
\begin{center}
\begin{tabular}{|c|c|} 
\hline
\rule[-1ex]{0pt}{3.5ex}  \bf Component\  & \bf Specifications\  \\
\hline\hline
\rule[-1ex]{0pt}{3.5ex}  \it Collimator \ & Field of view: 60mm diameter, Focal length :  755mm\\
\hline
\rule[-1ex]{0pt}{3.5ex}  \it Cross Disperser Prism \ & Apex angle : 40 degree, Base size: 126 x 165mm  \\
\hline
\rule[-1ex]{0pt}{3.5ex} \it Echelle grating \ & Grove density: 52.67 grooves/mm, Blaze angle 70 degree   \\
\hline
\hline
\end{tabular}
\end{center}
\end{table}
\begin{figure}[ht]
\begin{center}
\includegraphics[scale = 0.7]{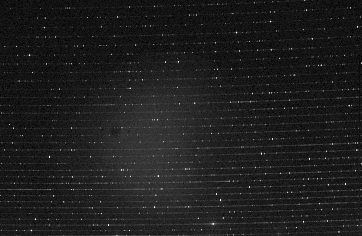}
\end{center}
\caption
{ \label{fig:THAR}
A raw image of the Th-Ar calibration spectra taken with an exposure time of 120~s. The analysis presented in this paper was done on  a set of 18 Th-Ar calibration frames taken between March 30, 2016 to April 2, 2016.}
\end{figure}

\section{Spectrograph Stability Analysis}
Originally, the spectrograph was designed as a general purpose instrument to meet high resolution and high throughput requirements along with the ease of operation \cite{VBT_3}. Since its inception in 2005, the instrument was extensively used for spectroscopic observations pertaining to abundances analysis, chemical composition, study of pulsating variables and investigation of stellar and galactic kinematics etc. For these science cases, the spectrograph stability is not very critical since relevant observations are made in a relatively short span of time. Therefore, moderate changes in the response of non-stabilized spectrographs is usually within the tolerance limits. However, instrumental instabilities becomes a major limitation for science programmes relying on high precision Doppler spectroscopy, spanning a long observing time base. In RV technique, a small Doppler shift in spectral lines has to be
measured  accurately.  For  our  spectrograph,  e.g.,  the
required precision of 100~m/s demands a detection of a 0.04-pixel pixel shift in the spectral lines of the star. Thus, it is essential to quantify the inherent stability of the spectrograph and also the accuracy of the algorithm used in such analysis.

The spectrograph is analysed for various stability issues
related to high precision studies. We examined the instrument induced shifts by measuring the deviation in the positions of the Th–Ar emission lines over a period of time. All observations and analysis related to this work were done with a slit width of 60~$\mu$m which corresponds to spectral resolution $R\approx$ 60,000. All Echelle orders within
the full wavelength range of the spectrograph were considered in the stability analysis. Figure \ref{fig:THAR} shows a central portion of one of the Th–Ar spectra obtained with the
spectrograph.
\begin{figure}[htb]
\begin{center}
\includegraphics[scale = 0.45]{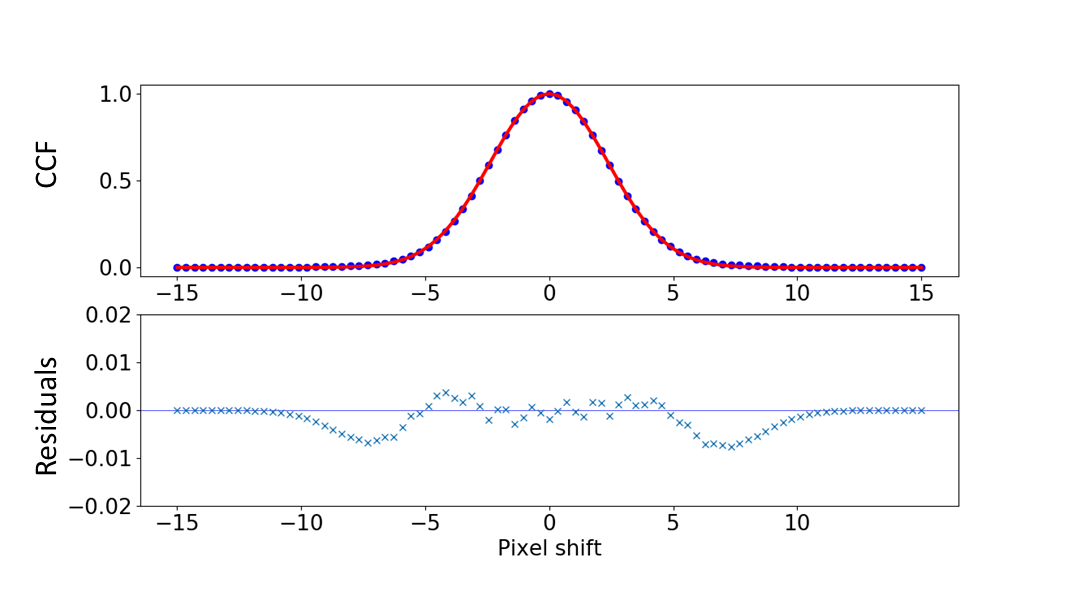}
\end{center}
\caption
{\label{fig:Fit}
A measured cross-correlation output of archival Th-Ar data from HARPS spectrograph. Gaussian fitting was done to locate the peak of the cross-correlation function.}
\end{figure}

\subsection{Procedure}
Multiple Th-Ar calibration frames were acquired with 4k$\times$4k spectrograph CCD camera between the nights of 30 March to 2 April 2016. We followed the standard data reduction procedures such as bad pixel removal, scattered light subtraction, bias corrections, aperture extraction and flat fielding. A dedicated data analysis routine was developed in Python for evaluating the stability of the line positions using cross-correlation algorithm \cite{Crosscorr_4}. The algorithm analyzes the shifts in the line positions by correlating the first Th-Ar calibration frame with rest of the spectra. The deviation in line positions from the reference spectra was obtained by fitting a Gaussian model to the output of the cross-correlation function (CCF). To reach sub-pixel accuracy we have interpolated the original data before applying the cross-correlation and extracted the peak using least-square Gaussian fit. This routine was customised to work on the individual spectral order, thereby inspecting the disturbances down to individual orders across the full Echelle frame. Our Th-Ar lamp has about 15 bright lines in red part (between order 30-49) of the spectrum causing pixel saturation. The presence of saturated lines produces a broad CCF profile whose peak cannot be determined accurately. Therefore, all strong lines were masked (ignored) in the final CCF analysis.
\begin{figure}[ht]
\begin{center}
\includegraphics[scale = 0.45]{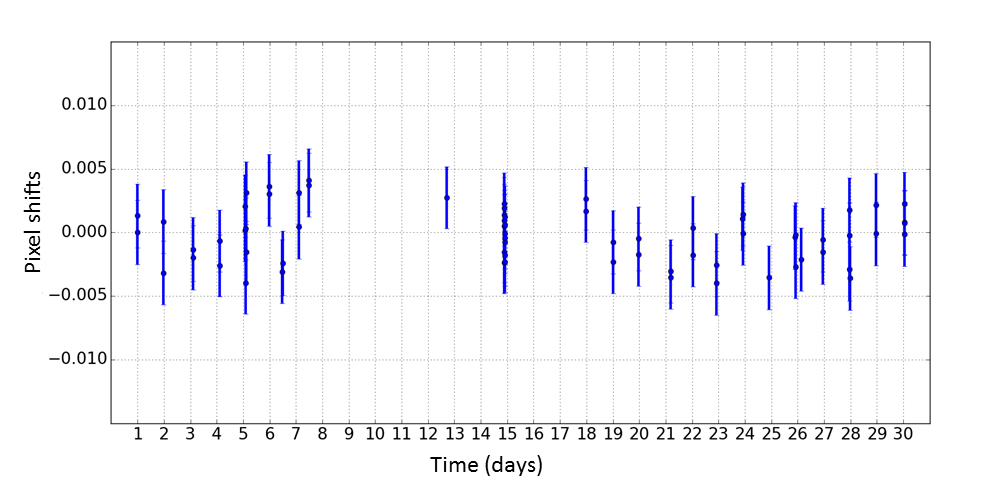}
\end{center}
\caption
{ \label{fig:Fit_HARPS}
Stability analysis on HARPS data using our data analysis routine. The stability of an unblended spectral line was analysed for data taken over one month. The RMS variation is seen confined to  0.001 pixel which is consistent with HARPS performance \cite{HARPSarchive_5}.}
\end{figure}

\subsection{Validation of the analysis algorithm}
HARPS is the fiber-fed spectrograph at 3.6 m telescope in La Silla \cite{HARPSintro_6}. This facility is used for high precision RV measurements with highest stability and accuracy available at present. The stability of the HARPS studied over a period of one month is discussed in : Ref \cite{HARPSpaper_7}. We obtained the same month long Th-Ar data from HARPS archive to test the accuracy of our analysis routine \cite{HARPSarchive_5}. Figure \ref{fig:Fit} shows a representative plot of cross-correlation output of HARPS data.

The stability analysis of HARPS was evaluated by measuring the position of unblended Th-Ar lines on the CCD. The absolute position of the line on the CCD in terms of pixels has been analysed over a period of 1 month during which the spectrograph stability was monitored. We performed the similar analysis on a single, well isolated high SNR Th-Ar line from the data to validate our algorithm. Figure \ref{fig:Fit_HARPS} shows the analysis output of our algorithm on HARPS data which is in good agreement with stability analysis reported in original study \cite{HARPSpaper_7}.
\begin{figure}[ht]
\begin{center}
\includegraphics[scale = 0.45]{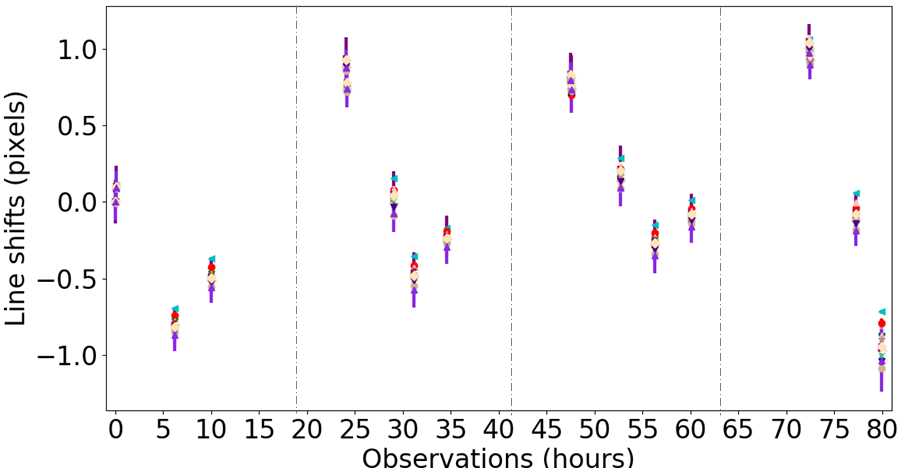}
\end{center}
\caption
{\label{fig:4days}
Measured line shift for VBT Echelle during 4 days observational test run. For clarity, each day's observation is separated by vertical dashed lines. The x-axis displays the time progression, in hours, from first observation to last. The y-axis displays the order-wise line shift computed for each consecutive observation. The coloured circles stacked along vertical direction represent shift for each Echelle order. The purple line shows the largest error (for order 15) in centroid detection.}
\end{figure}

\section{VBT Echelle Stability: Spatial/Spectral variations}
After validating our code on HARPS data, the stability analysis routine was tested on Th-Ar data obtained from VBT Echelle between 30 March to 2 April 2016. During the test run, we took 18 sets of Th-Ar observations, starting each evening to the next day morning, for 4 consecutive nights. On each night 3-4 calibration frames were obtained with separation interval about 3 hours between each frame. The first observational frame was selected as the reference frame for further analysis. The order-wise shift was obtained by cross-correlating individual Echelle order in each frame with the corresponding order in the reference frame.

Figure \ref{fig:4days} shows the results of stability analysis for the entire test run. For each observation, a shift in various Echelle orders is designated by coloured data points stacked along the vertical direction. Two broad trends are clearly seen in the Figure \ref{fig:4days}  a) The zero-point of the spectrograph showing large RMS drifts up to $0.8$~pixel between different observations and b) the intra-order $0.06$~pixel RMS scatter (vertical spread) between the reference frame and subsequent test frames.

We also note that zero-point shift is significantly high for early observations made on each night. We realised that these observations were made soon after the liquid nitrogen (LN) was filled in CCD dewar. The large systematic deviations in Figure \ref{fig:4days} can thus be attributed to the
fact that spectrograph did not have enough time to stabilise after the routine LN filling in evenings. The optimal utilisation of spectrograph for RV studies, therefore, requires that instrument is kept ready well in advance prior to observations. Next, we discuss the impact of environmental factors, namely temperature and pressure changes, on RV performance of a non-stabilized spectrograph.

\subsection{Effect of Temperature (T) and Pressure (P) Variations}
The prevailing ambient conditions $(T, P)$ impact the spectrograph stability in two major ways. First, the temperature excursions directly perturb the dimensional stability and alignment of the spectrograph at component levels. The spectral image can randomly wander on CCD plane due to non-zero thermal expansion coefficients (TEC) of the mechanical structure,  mounts, relay optics, prism, grating etc. Changes in grating spacing due to thermal expansion alone can contribute to velocity errors exceeding 100~m/s \cite{gratingtemp_8}.

\begin{figure}
\begin{center}
\includegraphics[scale = 0.55]{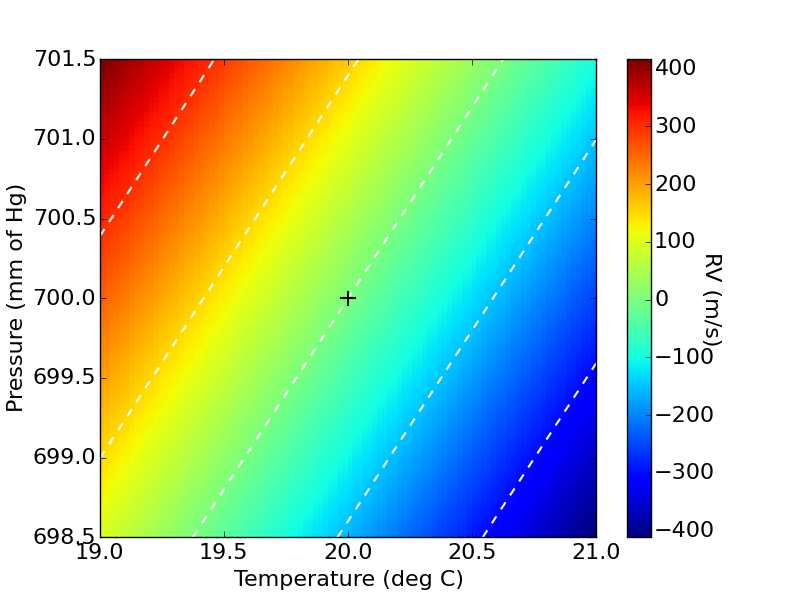}
\end{center}
\caption
{\label{fig:Temp_Press}
RV errors caused by variations of refractive index of air and its dependence on  pressure and temperature fluctuations. The cross symbol in the middle indicates the nominal operating conditions of $T\,\&\,P$ in the Echelle room. White dashed lines are tracers of same RV error in $T-P$ space. }
\end{figure}

Second, the changes in temperature and air pressure inside the spectrograph room also effect the ambient refractive index of  air. A standard dispersion formula for the dry air is given by \cite{Telluric_9}:
\begin{equation}
(n_{o}-1)= \left\{64.328 + \frac{29498.1}{146-\lambda^{-2}} + \frac{255.4}{41-\lambda^{-2}}\right\}\times10^{-6}
\end{equation}
where ($n_{o}-1$) is  air refractivity at standard temperature ($T=15^\circ$~C) and pressure ($P=760$~mm of Hg) and $\lambda$ is wavelength of light in $\mu$m. The dependence of dry air refractivity on temperature and pressure can be expressed as:
\begin{equation}
(n-1)=(n_{o}-1)\times\left\{ \frac{P[1+(1.049 - 0.0157T)\times 10^{-6}P]}{720.883(1 + 0.003661T}\right\}
\end{equation}
where $n$ is the refractive index at any other temperature $T$ (deg C) and pressure $P$ (mm of Hg). We can compute the refractive index change $\Delta n(T,P)$ from Eqs.~(2) and convert it to radial velocity changes as:
\begin{equation}
v=(n_2-n_1)\,c=\Delta n(T,P)\; c
\end{equation}
where $\Delta n$ is refractive index change due to pressure and temperature changes, $c$ is the light speed and $v$ is the RV error.  Figure \ref{fig:Temp_Press} shows the RV error computed using eq(1)-eq(3) on account of typical temperature ($T =20\pm1^{\circ}$) and pressure ($P =700 \pm1.5$ mm of Hg) fluctuations in the air conditioned VBT Echelle room. We note that departure from nominal operating condition can induce $\pm400$ m/s velocity swing in the spectral lines. Here, we have not attempted to model the effect of individual components of the spectrograph but the line shifts measured from the test observations are consistent with prediction based on eq(3).

\begin{figure}
\begin{center}
\begin{tabular}{c}
\includegraphics[scale = 0.5]{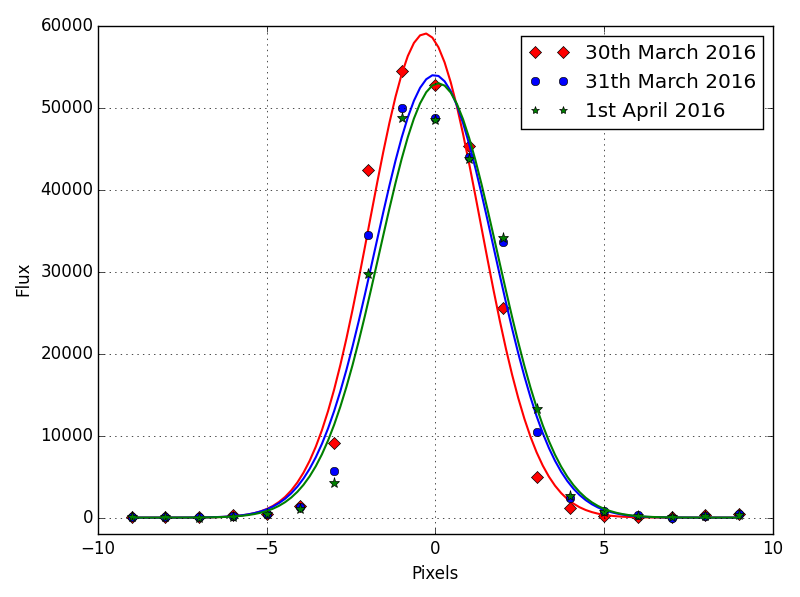}
\end{tabular}
\end{center}
\caption
{\label{fig:PSF_FIT}
Instrument profile (IP) sampled for an isolated Th-Ar Peak for three observations. The solid lines are single Gaussian fits to data shown by different colour markers.}
\end{figure}

Discrepancies in the model of the Instrument Profile (IP) and that of the observed spectra results in inaccuracies in RV measurements \cite{PSF_variations_10}. From Th-Ar spectra, we have also examined the IP and its variations with time and also across the spectral frame. We find a notable structure in IP which shows significant departure from a simple Gaussian or Lorentzian shape. For illustration, we show the IP in Figure \ref{fig:PSF_FIT}, extracted from an isolated Th-Ar line for three observations taken at different times. Imperfections in single Gaussian fit to data are clearly evident. The observed shape of the IP depends on time varying environmental perturbations and field dependent (static) aberrations in the optical design of the spectrographs. The main motivation of using an Iodine cell is to track the spectrograph's IP with simultaneous illumination. While developing the Iodine analysis code, we will model the IP with a multi-gaussian fit, as it is done normally.

\section{Scope for Improvement}

The main goal of the present analysis was to estimate the spectral image motion on the CCD detector plane of the spectrograph. This evaluation process has impressed upon the need to take remedial steps to mitigate the impact of noise sources \cite{improve_11}. To improve the baseline  stability of the instrument for achieving reasonable RV precision ($<100$ m/s)  with Iodine cell, following measures are planned as a part of instrument up-gradation:
\begin{itemize}
\item {\it Improved auto-guiding:} Errors in keeping the starlight focused at the centre of the fibre induces intractable shifts in the IP, resulting in RV inaccuracies.  An auto-guider system is being developed to eliminate the guiding errors in the spectrograph. The guiding assembly will be placed in front of the prime focus of the telescope, to guide the light into the optical fibre carrying the light to the spectrograph assembly.
\item {\it Mechanical stability and realignment:} The whole spectrograph assembly has to be re-aligned for correcting the inherent disturbances like tilt and other aberrations distorting the IP across CCD field. The mechanical stability of the spectrograph is also planned to be upgraded. The grating in the spectrograph assembly is a movable component.  The optical mounting of the grating assembly is to be upgraded to provide better locking at fixed position.
\item {\it Modification in calibration assembly:} The calibration lamp assembly is to be stabilised to eliminate errors due to the former. In order to evaluate the instrumental drift between the observations, a simultaneous Th-Ar exposure assembly is being planned. After the stabilisation of the calibration lamp assembly, the shifts in the line positions during the time of wavelength calibration and the actual observations can be analysed for the drift errors.
\item {\it Improvement in air-conditioning:} The current air conditioning system in the Echelle room is old and it cannot provide temperature stability better than $\pm 1^{\circ}$ C. Our goal is to achieve a  temperature regulation $\pm0.5^{\circ}$ C or better. To that end, a new AC unit, multiple temperature sensors spread across the room, and a better thermal insulation layer around the spectrograph table has been planned.
\end{itemize}

\section{Conclusions}

Instrument instability is a major limitation for science programmes relying on high precision Doppler spectroscopy. Modern spectrographs are designed to reach 1-10~m/s RV precision. They are operated in a highly stable vacuum environment. VBT Echelle is a general purpose spectrograph, which is is not designed for vacuum conditions. However, we are taking measures to enhance its RV precision using Iodine absorption cell and improving the thermal and mechanical stability of the spectrograph.

We conducted a series of experiments to measure the stability of the spectrograph. Our analysis shows that spectrograph has an overnight drift exceeding 0.8 pixels and 1 pixel drift over 4 days of observations. The nightly drift was seen to be systematically modulated in all observations. In addition, individual orders in each Th–Ar frame shows different scatter. An independent record of the temperature is currently unavailable but we strongly sus- pect the day to night temperature variations to be responsible for modulating the pixel drift. Laboratory tests are being done for upgrading the spectrograph components and improve the stability of the air conditioning system. Error budgets are to be estimated for required RV accuracy. After achieving the required stability of the spectrograph, the data analysis model for Iodine absorption cell will be developed and tested.
\begin{acknowledgements}
Authors would like to thank all the members of the technical and observing support staff of the VBT, Kavalur, India.

\end{acknowledgements}


\end{document}